# Temperature-dependent hardness of diamond-structured covalent materials

Xing Feng[1], Jianwei Xiao[1], Bin Wen[1,*] Jijun Zhao[2], Bo Xu[1,†], Yanbin Wang[3], Yongjun Tian[1]

[1]Center for High Pressure Science, State Key Laboratory of Metastable Materials Science and Technology, Yanshan University, Qinhuangdao 066004, China

[2]Key Laboratory of Materials Modification by Laser, Ion and Electron Beams (Ministry of Education), Dalian University of Technology, Dalian 116024, China

[3]Center for Advanced Radiation Sources, University of Chicago, Chicago, Illinois 60439, USA.

**Abstract** Understanding temperature-dependent hardness of covalent materials is not only of fundamental scientific interest, but also of crucial technical importance. Here we proposed a temperature-dependent hardness formula for diamond-structured covalent materials on the basis of the dislocation theory. Our results show that, at low temperature, the Vickers hardness is mainly controlled by Poisson's ratio and shear modulus, with the latter playing a dominant role. With increasing temperature, the governing mechanism for plastic deformation switches from shuffle-set dislocation control to glide-set dislocation control, and hardness drops steeply at elevated temperature. Moreover, an intrinsic parameter, $a^3G$, is revealed for diamond-structured covalent materials, which measures the resistance to softening at high temperature. This temperature-dependent hardness model shows remarkable agreement with experimental data of diamond-structured covalent materials. Current work not only sheds lights on the physical origin of hardness, but also provides a practical guide for superhard materials design.

**Keywords**：hardness; temperature dependent; dislocation; diamond-structured covalent materials; superhard materials

---

Authors to whom any correspondence should be addressed
* E-mail address: wenbin@ysu.edu.cn (Bin Wen) Tel: +86 13933969655
† E-mail address: bxu@ysu.edu.cn (Bo Xu)

Hardness is defined as the ability of a material to resist being dented or scratched by another material [1]. In the past decades, many endeavors have been dedicated to understand the origin of materials hardness and to estimate it. Several hardness models, both macroscopic [1-5] and microscopic [6-9], have been established for a wide range of materials with varying degrees of success. For example, by using Chen's formulas [4], a perfect correspondence between the calculated and experimental values of hardness can be achieved for a wide variety of crystalline materials as well as bulk metallic glasses.

These hardness models were established for the ambient conditions with temperature effect less considered. However, materials in practical applications usually are processed or operated at variable temperatures where their mechanical behaviors and performance are different from those at room temperature. A temperature-dependent hardness model is therefore highly expected. Experimentally, it is very challenging to carry out temperature-dependent hardness measurements due to the complexities in sample preparation plus maintaining a stable temperature during the measurement, which usually contribute to substantial errors in hardness values. Nonetheless, for diamond-structured covalent materials, an intensified softening behavior is clearly demonstrated with increasing temperature [1, 10-14]. Formulae with parameters fitted from the experimental data were built to understand this softening behavior [1, 15]. Still, the fundamental mechanisms (such as the dislocation and microstructure contributions, the effect of loading conditions, *etc*.) behind this temperature-dependent hardness are basically ignored, and the applicability to other materials systems are unknown. It is thus urgently needed to establish a temperature-dependent hardness model taking account of the physical mechanisms, for both scientific and technological reasons.

The difficulty in assessing hardness, especially with variable temperature, partially lies in the fact that the hardness determined by a specific measurement method is an engineering quantity, and cannot be evaluated directly from quantum mechanics [16]. During the hardness measurement, a plastic deformation must occur in the tested sample (such as the permanent impression dented by the indenter), which is correlated with the dislocations behaviors in the sample [1]. While the dislocation-governed plastic deformation have been widely investigated for metals [17-19], understanding the plastic deformation and consequentially the hardness of covalent materials on the basis of dislocation is still making progress [20]. For both classes of materials, the plastic deformation is closely related to dislocation glides on the slip systems as well as the microstructure. Our recent work on nanotwinned diamond suggests the room-temperature ultrahigh hardness can be ascribed to the dislocation behaviors and nanotwinned microstructure [20-22]. In general, the nucleation and propagation of dislocations in materials can be activated thermally and/or by applied force [23], resulting in a substantial plastic deformation for hardness evaluation. Therefore, the temperature effect on hardness can be tackled by considerations of the involved dislocations, along with other effects due to the microstructure and the loading conditions, *etc*. As an exemplification, we report in this work a temperature-dependent hardness formula based on the dislocation theory for diamond-structured covalent materials.

It is known that hardness and yield strength of a material both reflect the resistance to plastic deformation, with Tabor's law describing the correlation between hardness and yield strength [24]. Furthermore, the yield strength of a material is directly related to the corresponding critical resolved shear stress (CRSS), as revealed by Schmid's law [23]. Consequently, hardness can be evaluated by CRSS. Here CRSS is defined as minimum shear stress required for a dislocation to slip. To understand the temperature-dependent hardness of diamond-structured covalent materials (as shown in Fig. 1a), a temperature-dependent CRSS was firstly deduced on the basis of the dislocation theory. Due to the strong directional covalent bonds result in a large Peierls barrier when the dislocation moves in covalent materials, limiting that the dislocation line must move along some certain directions. These dislocation line in covalent materials have been observed to propagate primarily through kinks, or specifically through kink pairs [25], as schematically shown in Fig. 1b. Therefore, their CRSS can be obtained by simulating an energy path for a dislocation kink-pair nucleation and migration processing. A detail modeling process is described as follows.

The total energy ($W$) of a kink-pair as a function of kink-pair width ($x$) and applied shear stress ($\tau$) contains four terms: the kink formation energy ($2W_f$), kink migration energy ($W_m$), kink-pair interaction energy ($W_{int}$), and work done by the applied stress ($W_\tau$). On the basis of dislocation theory [23], $W$ can be expressed as

$$W(x,\tau) = 2W_f + W_m + W_{int} + W_\tau = \frac{A_1 G b^2 h}{2\pi} \ln\left(\frac{R}{r}\right) + W_m - \frac{A_2 G b^2 h^2}{8\pi x} - hbx\tau, \tag{1}$$

where $b$ is the magnitude of Burgers vector, $h$ the kink height, $G$ the shear modules, $R$ the integral range of linear elasticity theory, $r$ the radius of dislocation core, $A_1 = \cos^2\beta + \frac{\sin^2\beta}{1-\nu}$ and $A_2 = \frac{(1+\nu)\cos^2\beta + (1-2\nu)\sin^2\beta}{1-\nu}$ with $\nu$ being Poisson's ratio and $\beta$ the angle between Burgers vector and the dislocation line.

As shown in Fig. 1c, $W$ as a function of $x$ oscillates with the lattice periodicity along <110> direction. The envelope of the local maxima forms a curve (dashed line in Fig. 1c), and the maximum of the envelope can be considered as the activation energy of dislocation motion. Mathematically, the critical kink-pair width corresponding the activation energy, $x_c$, can be determined with the first derivative test [23]. Note $W_m$ contributing to the local variation (oscillation) can be ignored when considering the envelope maximum. As a result,

$$x_c = \left(\frac{A_2 hb}{8\pi} \frac{G}{\tau}\right)^{1/2}, \tag{2}$$

and the activation energy as a function of $\tau$ can then be determined as

$$W_c(\tau) = \frac{A_1 Gb^2 h}{2\pi} \ln\left(\frac{x_c}{r}\right) - (hb)^{3/2}\left(\frac{A_2 G\tau}{2\pi}\right)^{1/2}. \tag{3}$$

Here $R$ in Eq. (1) is set as $x_c$ for an approximation.

When considering the temperature effect, the dislocation motion is controlled by both applied shear stress and thermal activation [23]. For a given set of temperature $T$, applied stress $\tau$, and plastic strain rate $\dot{\varepsilon}$, the temperature-dependent CRSS can be written with a transcendental equation based on Eq. (3) and Orowan's relation [23] as following,

$$\tau_c^T = \frac{2\pi}{A_2 h^3 b^3 G}\left[\frac{A_1 Gb^2 h}{2\pi}\ln\left(\frac{x_c}{r}\right) - k_B T \ln\left(\frac{\rho_m b \lambda_b \nu_D}{\dot{\varepsilon}}\right)\right]^2, \tag{4}$$

where $k_B$ is the Boltzmann constant, $\rho_m$ the density of mobile dislocations, $\lambda_b$ the mean free path of dislocations slipping over obstacles, and $\nu_D$ the Debye frequency. By substituting Eq. (2) into Eq. (4), we have

$$\frac{\tau_c^T}{G} = C_1\left[C_2 \ln\left(C_3 \frac{G}{\tau_c^T}\right) - \frac{k_B T}{a^3 G}\ln\left(\frac{\rho_m b \lambda_b \nu_D}{\dot{\varepsilon}}\right)\right]^2, \tag{5}$$

where $C_1 = \frac{2\pi a^6}{h^3 b^3 A_2}$, $C_2 = \frac{hb^2 A_1}{4\pi a^3}$, and $C_3 = \frac{hbA_2}{8\pi r^2}$ are dimensionless constants only related to Poisson's ratio of the materials and the geometry of involved dislocation. $a$ is the lattice parameter.

By following Schmid's and Tabor's law, a temperature dependent Vickers hardness formula for covalent materials can be expressed as

$$H_T = n\tau_c^T, \tag{6}$$

where $n$ is product of pre-factors in Schmid's law and Tabor's law.

For diamond-structured covalent materials, their main dislocation slip systems are of {111}<110> type [26, 27]. Due to the two-interpenetrating face-centered crystal (fcc) sublattices, dislocation slips on {111} planes can occur at two different glide planes, *i.e.*, glide-set and shuffle-set glide planes (Fig. 1a). Usually, a 1/2<110> glide-set dislocation can be dissociated into two glide-set 1/6<112> partial dislocations. In contrast, a 1/2<110> shuffle-set dislocation would stay intact due to excessively high energy required for the dissociation. Meanwhile, the large Peierls barrier with deep trough along <110> directions in diamond-structured materials limits the dislocation lines along these directions [27, 28]. Therefore, two types of dislocations, namely the 1/6<112> glide-set 90° partial (edge) dislocations and 1/2<110> shuffle-set perfect (screw) dislocations, dominate the plastic deformation in diamond-structured covalent materials, as indicated in Fig. S1 of Supplemental Information (SI).

The calculated activation energies as a function of applied shear stress for 1/6<112> glide-set 90° partial dislocation and 1/2<110> shuffle-set perfect dislocation based on Eq. (3) are shown in Fig. 1d, where a crossover is clearly revealed with increasing shear stress, indicating a competition of deformation mechanism between glide-set and shuffle-set dislocations. This result based on the dislocation theory is further confirmed by the molecular dynamics (MD) simulation for diamond (inset of Fig. 1d, also see SI part I for MD calculation detail), and is consistent with previous experimental observations [10].

Table 1 lists the information for 1/6<112> glide-set 90° dislocation and 1/2<110> shuffle-set 0° perfect dislocation in diamond-structured covalent materials. Others parameters, such as temperature-dependent lattice constant, shear modulus, and Poisson's ratio, *etc.*, were determined with the methods presented in Refs. [29-31] (see SI part II for calculation details). These parameters at 0 K are listed in Table 2. The temperature-dependent CRSS, $\tau_{c,s}^T$ for shuffle-set dislocation and $\tau_{c,g}^T$ for glide-set one, can then be evaluated from Eq. (5) with a geometric or numerical method. These two types of dislocations compete with each other, and the one with lower CRSS dominates the deformation at given temperature and shear stress. Therefore, the CRSS of the investigated materials can be determined as

$$\hat{\tau}_c^T = \min(\tau_{c,s}^T, \tau_{c,g}^T).  \quad (7)$$

For diamond-structured polycrystalline materials, their pre-factors in Schmid's law and Tabor's law are 3.1 [32, 33] and 2.74 [24], respectively. Therefore, Eq. (6) for diamond-structured covalent materials can be written as $H_T = 8.5 \hat{\tau}_c^T$.

Fig. 2a displayed the calculated Vickers hardness (at 300 K) for typical diamond-structured covalent materials compared with the experimental data [4], exhibiting a nice consistency. The temperature-dependent Vickers hardness of diamond, Si and Ge are calculated and plotted in Fig. 2b–d (see Fig. S2 for the temperature-dependent lattice constant, shear moduli, and Poisson's ratio), respectively, in good accordance with the experimental hardness values over a wide temperature range [1, 11-14]. In addition, the temperature-dependent Vickers hardness of the polar covalent materials of cubic BN and SiC are plotted in Fig. S3, which are also in good accordance with the experimental values. Moreover, the calculated transition temperatures from the shuffle-set dislocation controlled deformation to glide-set dislocation controlled one ($T_{s-g}$) are 1402.6 K, 676.8 K and 560.2 K for diamond, Si and Ge, respectively, which are comparable with the experimental values, *i.e.*, 1450 K for diamond [10], 650 K for Si [1], and 600 K for Ge [14].

At low temperature, the dislocation motions due to thermal activation can be ignored. In this case, the Vickers hardness determined from Eqs. (5)–(7) for a diamond-structured covalent material is determined by the shear modulus and Poisson's ratio, and Eq. (6) can be simplified as

$$H_0 = k_0(\nu)G, \tag{8}$$

where $k_0(\nu)$ is a proportional coefficient related to Poisson's ratio. Fig. 3 shows the Vickers hardness map at 0 K constructed from Eq. (6). Obviously, both shear modulus and Poisson's ratio contributes to the Vickers hardness. As shown in Fig. 3b, the Vickers hardness increase linearly with increasing shear modulus, and $k_0(\nu)$ decreases slightly with increasing Poisson's ratio. Compared with shear modulus, Poisson's ratio shows a less prominent effect on the Vickers hardness (Fig. 3c). It is clear that high shear modulus and low Poisson's ratio are essential for (super)hard materials. In addition, we note that the proportional coefficient $k_0(\nu)$ varies in a narrow range of 0.14–0.19 for $\nu$ in the range of 0–0.3 (see the inset of Fig. 3a). By fitting $k_0(\nu)$ with $\nu$, Eq. 8 can be rewritten as

$$H_0 = (0.18 + 0.05\nu - 0.51\nu^2)G. \tag{9}$$

Previously, some semi-empirical hardness models relating hardness to materials' shear modulus directly with proportional coefficients of 0.12 [1], 0.151 [3], 0.147 [5] fitted from experimental data. These values agree nicely with $k_0(\nu)$ calculated in current work, verifying the effectiveness of our consideration of hardness on the basis of dislocation dynamics.

At elevated temperature, the thermal activated dislocation motions start to function. From Eq. (5), $\tau_c^T/G$ is clearly correlated with Poisson's ratio, temperature, and a materials-related parameter of $a^3G$, which combined with Eq. (6) gives the temperature dependent Vickers hardness as

$$H = k(T, a^3G, \nu)G, \tag{10}$$

with the proportional coefficient $k$ as a function of $\nu$, $T$ and $a^3G$. The thermal and material's non-elastic properties effects are essentially included in $k$. Note that $k = H/G$ can be considered as a normalized hardness with respect to material's shear modulus.

Fig. 4 shows $k$ as a function of $\nu$, $T$ and $a^3G$. Considering the general range of $\rho_m$, $\lambda_b$, and $\dot{\varepsilon}$ for interested materials (such as those listed in Table 1), we find the estimated values of the second logarithm in Eq. 5, $\ln\left(\frac{\rho_m b \lambda_b v_D}{\dot{\varepsilon}}\right)$, vary in a relatively narrow range for different diamond-structured covalent materials (Fig. S4). Therefore, fixed values of 18.8 and 32.1 are used in the calculation for shuffle-set and glide-set dislocations, respectively. Similar simplification was previously used [34]. As shown in Fig. 4a, $k$ decreases with increasing temperature for a given set of Poisson's ratio and $a^3G$, indicating softening occurs at high temperature. Furthermore, for a fixed Poisson's ratio, the larger $a^3G$ is, the slower $k$ decreases, indicating that materials with larger $a^3G$ are more difficult to soften with increasing temperature. The transition temperature $T_{s-g}$ from shuffle-set dislocation control to glide-set dislocation control is calculated and shown in Fig. 4b. $T_{s-g}$ increases linearly with $a^3G$ for a given Poisson's ratio, meanwhile it goes to higher temperature with larger Poisson's ratio. It is justified to identify $a^3G$ as an intrinsic index for diamond-structured covalent materials, measuring the resistance to soften at elevated temperature, which is consistent with previous semi-empirical results [35].

Besides the above-mentioned intrinsic properties that determine hardness of materials, other factors, such as dislocation characteristic, microstructure and loading conditions of the sample, *etc.*, also show great impact on the hardness. These additional effects can easily be accounted for in current dislocation-based hardness model (Fig. S5). For example, the hardness varies with the density of mobile dislocations: the larger the dislocation density, the lower the hardness (Fig. S5a). The effects of loading speed (Fig. S5b) and grain boundary (Fig. S5c) on hardness have also been investigated.

In summary, a temperature-dependent Vickers hardness model has been developed on the basis of dislocation theory for diamond-structured covalent materials. At low temperature, the Vickers hardness is mainly controlled by Poisson's ratio and shear modulus with the latter playing a dominant role. At elevated temperature, the deformation mechanism changes from shuffle-set dislocation control to glide-set dislocation control, and the Vickers hardness is further affected by temperature and a material-related parameter of $a^3G$. Materials with larger $a^3G$ are more difficult to soften at elevated temperature. These findings help to unveil the physics of hardness, and can provide a direct guidance for superhard materials design, especially at high temperature.

**Acknowledgment**

This work was supported by the National Natural Science Foundation of China (Grant Nos. 51925105, 51771165, and 51525205), National Magnetic Confinement Fusion Energy Research Project of China (2015GB118001) and NSF (Grant No. EAR-1361276) and the National Key R&D Program of China (YS2018YFA070119).

**Table 1.** Geometric parameters for $\frac{1}{6}<112>$ glide-set 90° partial and $\frac{1}{2}<110>$ shuffle-set perfect dislocations.

| Dislocation type | $\beta$ | $b$ | $h$ | $r$ | $\rho_m$ (m$^{-2}$) | $\lambda_b$ (nm) | $\dot{\varepsilon}$ (s$^{-1}$) |
|---|---|---|---|---|---|---|---|
| $\frac{1}{2}<110>$ shuffle-set | 0° | $\sqrt{2}a/2$ | $\sqrt{6}a/4$ | $0.9b$ | $0.3\times10^8$ | 100 | $10^{-4}$ |
| $\frac{1}{6}<112>$ glide-set | 90° | $\sqrt{6}a/6$ | $\sqrt{6}a/4$ | $0.3b$ | $0.3\times10^{14}$ | 100 | $10^{-4}$ |

**Table 2.** Calculated lattice constants, shear moduli, $a^3G$, Poisson's ratios, Debye frequencies at 0 K and shuffle-set to glide-set transition temperatures for selected diamond-structured covalent materials. Hardness values are given for temperatures (0 K and 300 K) in comparison with experimental ones.

| Phase | $a$ (Å) | $G$ (GPa) | $a^3G$ (×10$^{-18}$ J) | $\nu$ | $\nu_D$ (THz) | $H_0$ (GPa) | $H_{300\,K}$ (GPa) | $H_{exp}$ (GPa) | $T_{s-g}$ (K) |
|---|---|---|---|---|---|---|---|---|---|
| Diamond | 3.57 | 521 | 23.7 | 0.07 | 39.3 | 94.7 | 85.6 | 60-150[a] | 1402.6 |
| Si | 5.47 | 64.2 | 10.5 | 0.21 | 15.1 | 10.9 | 8.9 | 11.3[b] | 676.8 |
| Ge | 5.78 | 45 | 8.7 | 0.19 | 8.3 | 7.8 | 6.1 | 7.2[c] | 560.2 |
| AlAs | 5.73 | 43.86 | 8.3 | 0.23 | 11.3 | 7.3 | 5.6 | 5[d] | 542.8 |
| AlP | 5.51 | 52 | 8.7 | 0.24 | 12.9 | 10.7 | 8.9 | 9.4[d] | 722.5 |
| AlSb | 6.23 | 32 | 7.7 | 0.23 | 9.8 | 5.3 | 4 | 4[d] | 509.7 |
| BAs | 4.82 | 129.4 | 14.5 | 0.13 | 21.4 | 23.1 | 19.8 | 19[b] | 886.4 |
| BN | 3.63 | 390 | 18.7 | 0.11 | 34.2 | 70.2 | 62.0 | 46-80[a] | 1125.5 |
| BP | 4.55 | 168 | 15.8 | 0.11 | 24.0 | 30.2 | 26.1 | 31[a] | 957.9 |
| GaAs | 5.76 | 43.68 | 8.3 | 0.22 | 7.7 | 7.3 | 5.8 | 7.5[b] | 550.4 |
| GaP | 5.53 | 54 | 9.1 | 0.22 | 10.8 | 9.1 | 7.2 | 9.5[d] | 597.4 |
| GaSb | 6.22 | 32.2 | 7.7 | 0.21 | 6.4 | 6.7 | 5.4 | 4.5[d] | 625.7 |
| InAs | 6.21 | 28.3 | 6.8 | 0.26 | 6.2 | 4.5 | 3.3 | 3.8[d] | 463.6 |
| InP | 6 | 34.1 | 7.4 | 0.26 | 9.4 | 5.5 | 4.1 | 5.4[d] | 498.7 |
| InSb | 6.65 | 22.28 | 6.6 | 0.25 | 5.0 | 3.6 | 2.7 | 2.2[d] | 445.9 |
| SiC | 4.38 | 198.74 | 16.7 | 0.14 | 24.1 | 33.4 | 29.4 | 26-37[a] | 1076.4 |

[a] Vickers hardness from Ref. [36]
[b] Knoop hardness from Ref. [37]
[c] Vickers hardness from Ref. [38]
[d] Knoop hardness from Ref. [39]

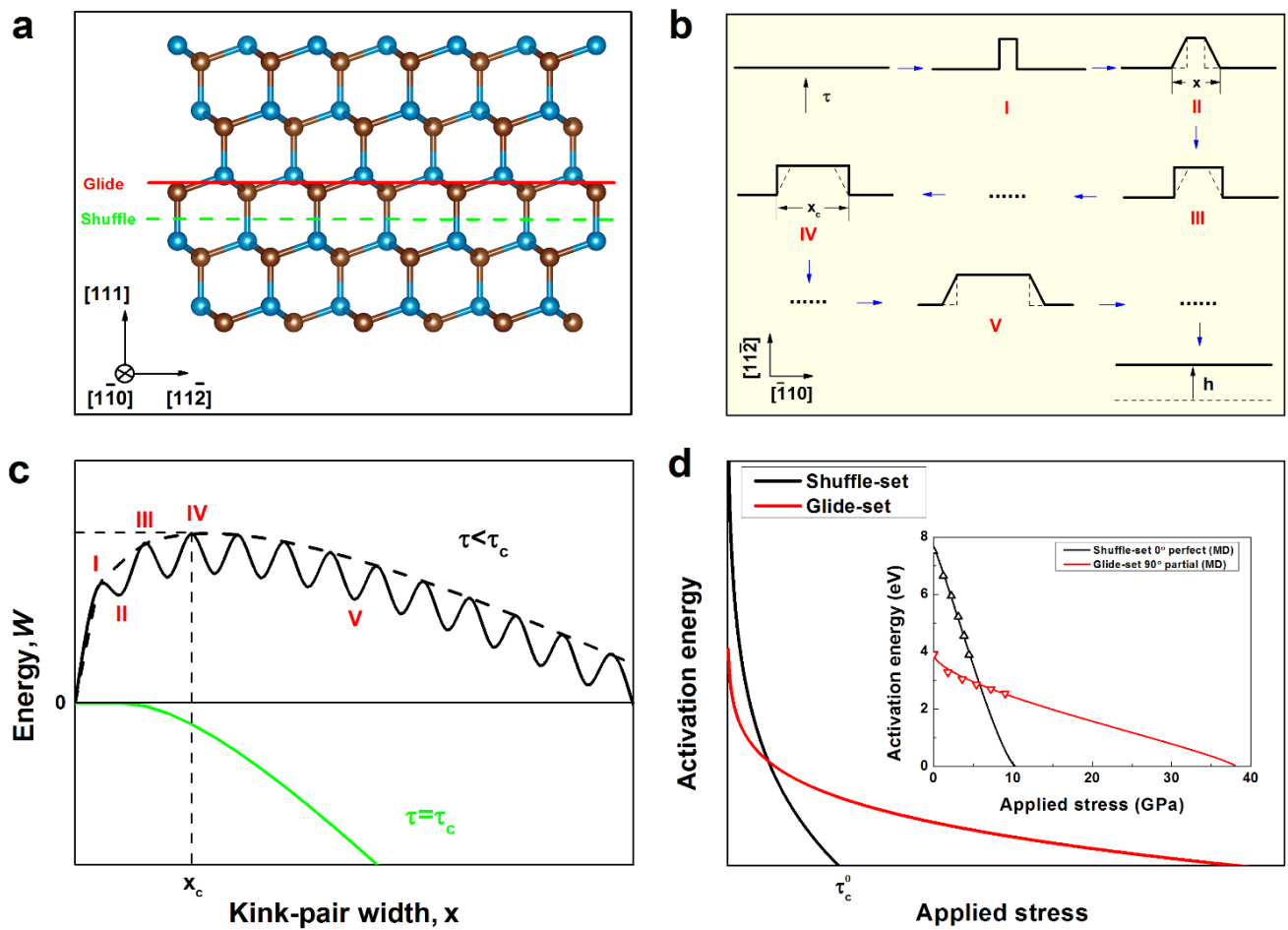

**Fig. 1. Schematic diagram for the computational method used in this study.** (a) {110} projection of the diamond-structured lattice. The green and red lines indicate the {111} shuffle and glide planes, respectively. (b) Kink-pair nucleation and motion process under applies stress. Shear stress $\tau$ acting perpendicularly to a dislocation line parallel to $<\bar{1}10>$ produces a kink pair (I), which expands subsequently (II through V), resulting in an upward motion of the dislocation line in $<11\bar{2}>$ direction with a step of *h*. (c) Total energy variation with respect to kink-pair width under different applied shear stress conditions. The oscillation reflects lattice periodicity. (d) Activation energy as a function of applied stress for shuffle-set and glide-set dislocation motion. The inset shows the MD result for diamond.

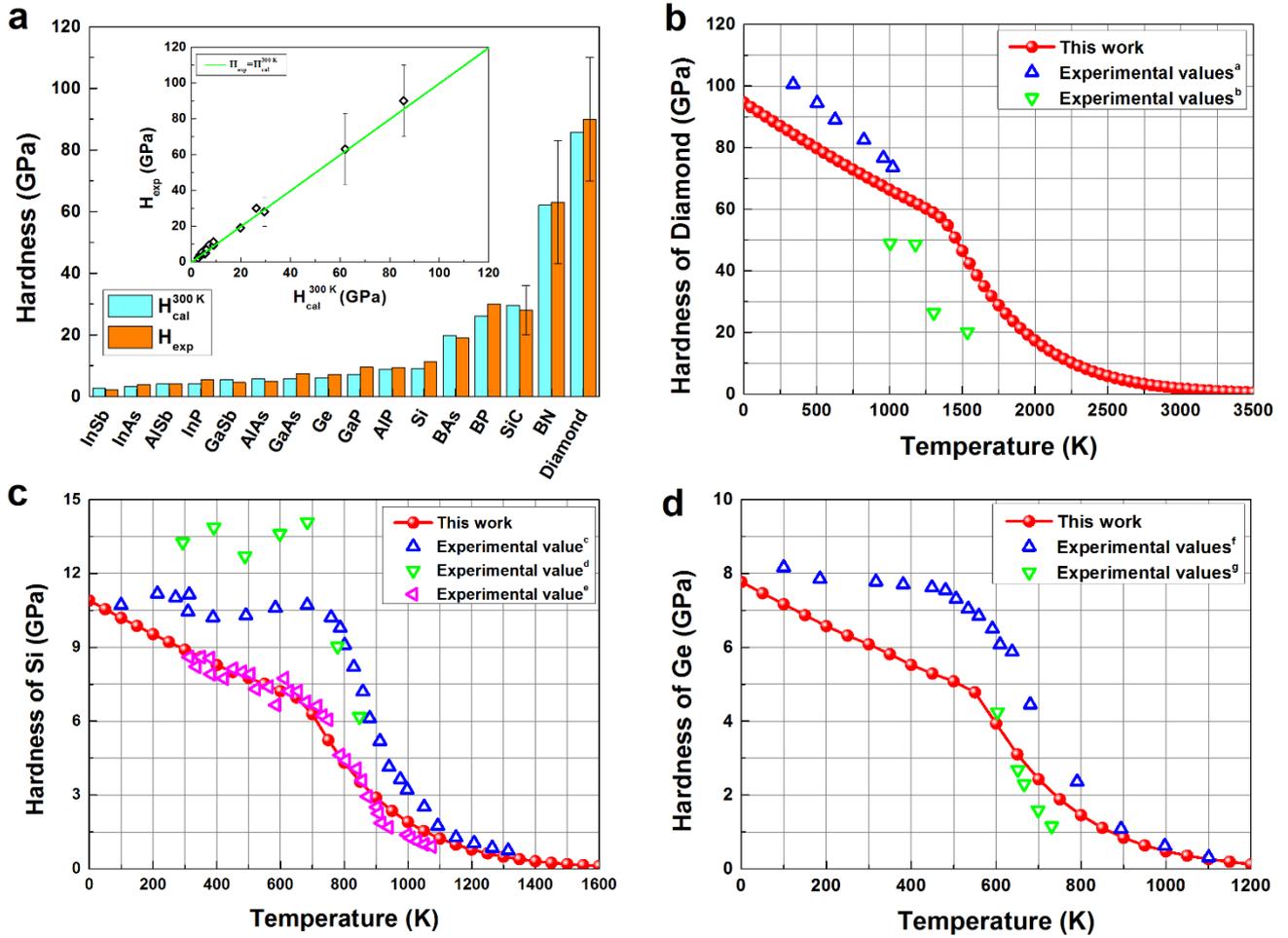

Fig. 2. Calculated Vickers hardness compared with experimental values. (a) Comparison of Vickers hardness values from current work and experimental results at 300 K. (b–d) Calculated temperature dependent Vickers hardness for diamond, Si, and Ge in comparison with experimental data ([a]Ref. [11], [b]Ref. [10], [c]Ref. [12], [d]Ref. [13], [e]Ref. [1], [f]Ref. [14], [g]Ref. [13]).

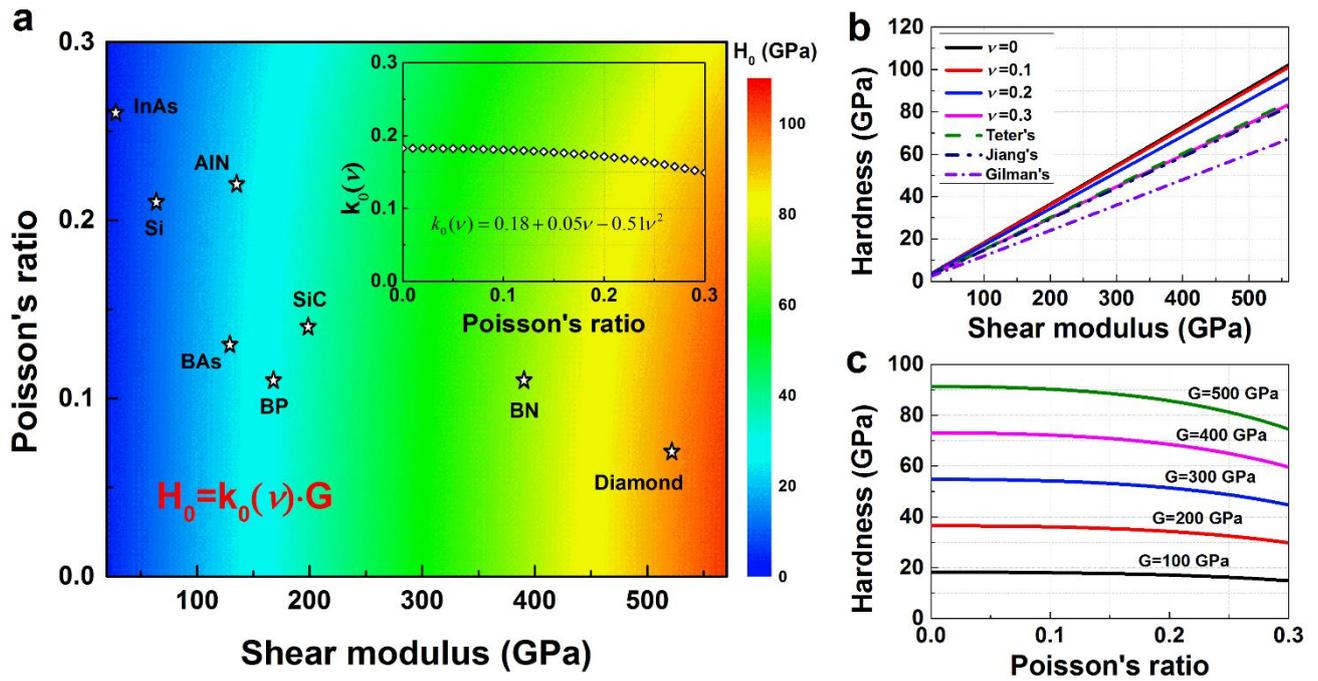

**Fig. 3. Effect of shear modulus and Poisson's ratio on hardness of diamond-structured covalent materials at 0 K.** (a) Calculated Vickers hardness map as a function of shear modulus and Poisson's ratio. (b) Effect of shear modulus on materials hardness with different Poisson's ratio. (c) Effect of Poisson's ratio on materials hardness with different shear modulus.

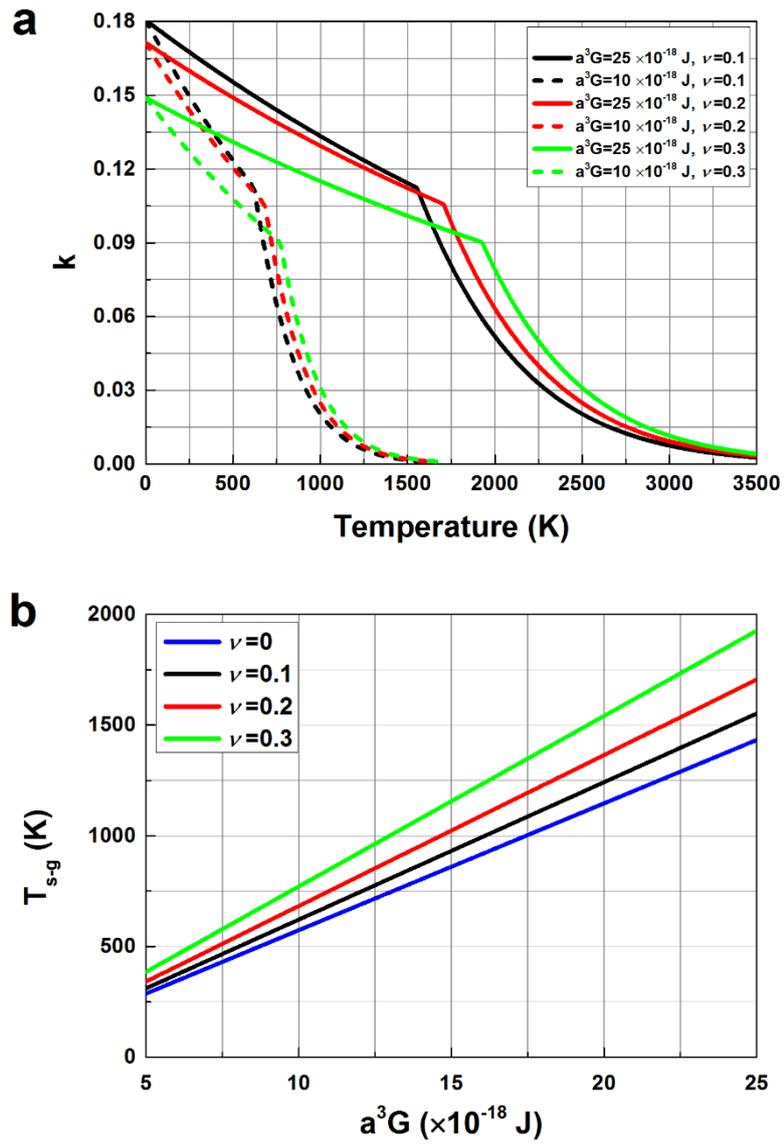

**Fig. 4. Temperature effect on Vickers hardness of diamond-structured covalent materials.** (a) The effects of temperature, $a^3G$, and Poisson's ratio on $k$. (b) The effect of $a^3G$ and Poisson's ratio on the transition temperature $T_{s-g}$.